\documentclass{PoS}
\usepackage{cite,epsfig}

\title{
\vspace*{-1.9cm}
\begin{minipage}{\textwidth}
{\normalfont\small LTH 1150, DESY 18-010, Nikhef 2018-004
\hspace{\fill} January 2018}\\
\end{minipage}\\[20pt]
 Four-loop results on anomalous dimensions and splitting functions in QCD}

\ShortTitle{Four-loop results on anomalous dimensions and splitting functions}

\author{\speaker{A. Vogt}\\
        \mbox{Department of Mathematical Sciences, University of Liverpool,
        Liverpool L69 3BX, UK}\\
        E-mail: \email{Andreas.Vogt@liverpool.ac.uk}
        }

\author{S. Moch \phantom {g} \\
       II.~Institut f\"ur Theoretische Physik, Universit\"at Hamburg,
       D-22761 Hamburg, Germany\\
       E-mail: \email{sven-olaf.moch@desy.de}
       }

\author{B. Ruijl \phantom {g} \\
       Institute for Theoretical Physics, ETH Z\"urich, 8093 Z\"urich, 
       Switzerland\\
       E-mail: \email{bruijl@phys.ethz.ch}
       }

\author{T. Ueda, J.A.M. Vermaseren \phantom {g} \\
       \mbox{Nikhef Theory Group, Science Park 105, 1098 XG Amsterdam,
       The Netherlands}\\
       E-mails: \email{tueda@nikhef.nl, t68@nikhef.nl}
       \\ \\ \\ }

\abstract{
We report on recent progress on the flavour non-singlet splitting functions in
perturbative QCD. The~exact four-loop (N$^3$LO) contribution to these functions
has been obtained in the planar limit of a large number of colours.
Phenomenologically sufficient approximate expressions have been obtained for
the parts not exactly known so far.  Both cases include results for the
four-loop cusp and virtual anomalous dimensions which are relevant well beyond
the evolution of non-singlet quark distributions, for which an accuracy of
(well) below 1\% has now been been reached.
}

\FullConference{13th International Symposium on Radiative Corrections 
(Applications of Quantum Field Theory to Phenomenology)\\
25-29 September 2017, St. Gilgen, Austria}

\usepackage{array}
\newcolumntype{T}{>{\ttfamily} c}
\newcolumntype{M}{>{$\displaystyle} c <{$}}

\newcommand{\beq}{\begin{equation}}
\newcommand{\eeq}{\end{equation}}
\newcommand{\bea}{\begin{eqnarray}}
\newcommand{\eea}{\end{eqnarray}}
\newcommand{\nn}{\nonumber}
\newcommand{\nin}{\noindent}

\newcommand{\hspp}{{\hspace{4mm}}}
\newcommand{\hspn}{{\hspace{-4mm}}}

\def\frct#1#2{\mbox{\small{$\displaystyle\frac{#1}{#2}$}}}

\newcommand{\MSb}{$\overline{\mbox{MS}}$}
\newcommand{\als}{\alpha_{\sf s}}
\newcommand{\ars}{a_{\sf s}}
\def\ar(#1){{a_{\sf s}^{\,#1}}}
\def\as(#1){{a_{\sf s}^{\,#1}}}
\def\z#1{{\zeta_{\:\!#1}^{}}}

\def\zts{\zeta_{\:\!3}^{\,2}}

\def\xm1{{(1 \! - \! x)}}

\def\nc{{n_c}}
\def\ncs{{n_{c}^{\,2}}}
\def\nct{{n_{c}^{\,3}}}

\def\ca{{C_{\!A}}}
\def\cas{{C^{\, 2}_{\!A}}}
\def\cat{{C^{\, 3}_{\!A}}}

\def\nc{{n^{}_{\! c}}}
\def\nfz{{n^{\,0}_{\! f}}}
\def\nf{{n^{}_{\! f}}}
\def\nfo{{n^{\,1}_{\! f}}}
\def\nfs{{n^{\,2}_{\! f}}}
\def\nft{{n^{\,3}_{\! f}}}
\def\cf{{C_{\! F}^{}}}
\def\cfs{{C^{\, 2}_{\! F}}}
\def\cft{{C^{\, 3}_{\! F}}}
\def\cff{{C^{\, 4}_{\! F}}}

\def\dfFAnc{{d_F^{\,abcd}d_A^{\,abcd}/{N_R }}}

\begin{document}

\section{Introduction}
\vspace*{-1mm}

\nin
Up to power corrections, observables in {\it ep} and {\it pp} hard scattering 
can be schematically expressed~as
\bea 
\label{OepOpp}
  O^{\,ep} \;=\; f_{i}^{} \,\otimes\, c_i^{\,\rm o} \; , \quad
  O^{\,pp} \;=\; f_{i}^{} \,\otimes\, f_{k}^{} \,\otimes\, c_{ik}^{\,\rm o}
\eea
in terms of the respective partonic cross sections (coefficient functions)
$c^{\,\rm o}$ and the universal parton distribution functions (PDFs)
$f_{i\,}^{}(x,\mu^2)$ of the proton at a (renormalization and factorization) 
scale $\mu$ of the order of a physical hard scale, e.g., $M_H$ for the 
total cross section for the production of the Higgs boson.  
The dependence of the PDFs on the momentum fraction $x$ is not calculable in 
perturbative QCD; their scale dependence is governed by the 
renormalization-group evolution~equations
\beq
\label{Evol}
 \frct{\partial}{\partial \ln \mu^2} \, f_i^{}(x,\mu^2) \: =\: 
 \left[ P^{}_{ik}(\als(\mu^2)) \otimes f_k^{}(\mu^2) 
 \right]\!(x)
\eeq
where $\otimes$ denotes the Mellin convolution.
The splitting functions, which are closely related to the anomalous dimensions
of twist-2 operators in the light-cone operator-product expansion (OPE), and 
the coefficient functions can be expanded in powers of the strong coupling
$\ars \equiv \als(\mu^2)/(4\pi)$, 
\bea
\label{Pexp}
 P \, & \:=\: & \quad
    \ar()\, P^{\,(0)} 
       +\, \ar(2)\, P^{\,(1)}
       +\, \ar(3)\, P^{\,(2)} 
     \,+\, \ar(4)\, P^{\,(3)} 
       +\: \ldots 
\;\;\; , \\
\label{Cexp}
 c^{\,\rm o}_{a} & \:=\: & \ars^{\:n_{\rm o}}
   \big[ \, c_{\rm o}^{\,(0)} 
   \;+\: \ar()\, c_{\rm o}^{\,(1)} 
   \;+\: {\ar(2)\, c_{\rm o}^{\,(2)}} \,
   \:+\: {\ar(3)\, c_{\rm o}^{\,(3)}} 
   \:+\: \ldots \,\big]
\; .
\eea
Together the first three terms in eqs.~(\ref{Pexp}) and (\ref{Cexp}) provide
the next-to-next-to-leading order (N$^{2}$LO) of perturbative QCD for the 
observables (\ref{OepOpp}). 
This is now the standard approximation for many hard processes; see 
refs.~\cite{P2ns,P2sg,DP2,DP2v} for the corresponding splitting functions.

\vspace*{1mm}
Corrections beyond N$^{2}$LO are of phenomenological interest 
where high precision is required, such as in determinations of $\als$ from 
deep-inelastic scattering (DIS) (see refs.~\cite{mvvC2L,mvvC3} for the 
N$^3$LO corrections to the most important structure functions), 
and where the perturbation series shows a slow convergence, such as for Higgs 
production via gluon-gluon fusion calculated in ref.~\cite{N3LOggH} at N$^3$LO.
The size and structure of the corrections beyond N$^2$LO are also of 
theoretical interest.

\vspace*{1mm}
Here we briefly report about considerable recent progress on the three 
four-loop (N$^3$LO) non-singlet splitting functions. We focus on the 
quantities $P_{\,\rm ns}^{\,\pm(3)}(x)$ for the evolution of
flavour-differences
  $ q_i^{} \pm \bar{q}_i^{} - (q_k^{} \pm \bar{q}_k^{}) $ 
of quark and antiquark distributions; for more details see ref.~\cite{MRUVV}.

\vspace*{-1mm}
\section{Diagram calculations of fixed-$N$ moments}
\vspace*{-1mm}

\nin
Two methods have been applied for obtaining Mellin moments of the quantities 
$P^{\,(3)}$ in eq.~(\ref{Pexp}). Depending on the function, both can be used 
to determine the same even-$N$ {\it or} the odd-$N$ moments.
 
\vspace*{1mm}
In the first one calculates, via the optical theorem and a dispersion relation 
in $x$, the unfactorized structure functions in DIS, as done at two and three 
loops in refs.~\cite{LV1991,3loopN1,3loopN2,3loopN3}.
The construction of the {\sc Forcer} program \cite{Forcer} has facilitated the
extension of those computations (which also provide moments of the coefficient 
functions) to four loops. For the hardest diagrams, the complexity of these 
computations rises quickly with $N$, hence  only $N \leq 6\,$ has been covered 
completely so far \cite{avLL2016}. 
Much higher $N$ can be accessed for simpler cases, e.g., values up to $N\!>40$
have been reached for high-$\nf$ parts. These were sufficient to determine 
the complete $\nfs$ and $\nft$ parts of the non-singlet splitting functions
$P_{\,\rm ns}^{\,(3)}(x)$ and the $\nft$ parts of the corresponding 
flavour-singlet quantities \cite{DRUVV}.

The increase of the complexity of the Feynman integrals with $N$ is more 
benign for the second method based on the OPE which was applied  to the 
present non-singlet cases at NLO in ref.~\cite{FRSns77}, see also 
ref.~\cite{BBK09ope}.
{\sc Forcer} calculations in this framework have reached $N=16$ for all
contributions to the functions $P_{\,\rm ns}^{\,(3)}$, $N=18$ for their
$\nf$ parts and $N=20$ for the complete limit of a large number of colours
$\nc$ \cite{MRUVV}.
See refs.~\cite{P3BC1,P3BC2,VelizN2,VelizN34} for earlier calculations 
of $P_{\,\rm ns}^{\,\pm (3)}$ at $N \!\leq 4$.

\vspace*{-1mm}
\section{Towards all-{\large $N$} expressions}
\vspace*{-1mm}

\nin
If the anomalous dimensions 
   $\gamma_{\,\rm ns}(N) = -\,P_{\,\rm ns}(N)$ 
at N$^{\,n>2}$LO are analogous to the lower orders, then they can be expressed 
in terms of harmonic sums $S_{\vec{w}}$ \cite{HSums1,HSums2} and denominators 
$D_a^{\:k} \equiv (N\!+\!a)^{-k}$ as
\beq
\label{gForm}
  \gamma_{\,\rm ns}^{\,(n)}(N) \;=\;
  \sum_{w=0}^{2n+1} c_{00\vec{w}}^{}\, S_{\:\!\vec{w}}(N) 
  \,+\,
  \sum_{\;a^{\phantom{a}}} \, \sum_{k=1}^{2n+1} \, \sum_{w=0}^{2n+1-k} \!
  c_{ak\vec{w}}^{}\, D_a^{\:k}\, S_{\:\!\vec{w}}(N) 
\; . 
\eeq
The denominators at the calculated values of $N$ indicate $a=0,1$ for 
$\gamma_{\,\rm ns}^{\,\pm\,}$, with coefficients $c_{00\vec{w}}$, 
$c_{ak\vec{w}}$ that are integer modulo low powers of 1/2 and 1/3.
Sums up to weight $w=2n+1$ occur at N$^{\,n}$LO.
 
\vspace*{1mm}
Based on a conformal symmetry of QCD at an unphysical number of 
space-time dimensions~$D$, it has been conjectured that the \MSb\ functions 
$\gamma_{\,\rm ns}^{}(N)$ are constrained by `self-tuning'
\cite{BassoK06,DokMar06},
\beq
\label{Stune}
  \gamma_{\,\rm ns}^{}(N) \;=\; \gamma_{\,\rm u}^{}
  \left( N + \sigma \,\gamma_{\,\rm ns}^{}(N)-\beta(\ars)/\ars ) \right)
\eeq
where $\beta(\ars) = - \beta_0 \,\ar(2) - \beta_1 \,\ar(3) - \ldots $ is the 
beta function, for its present status see \mbox{refs.~\cite{beta4a,beta4b}}.
The initial-state (PDF) and final-state (fragmentation-function) anomalous
dimensions are obtained for $\sigma=-1$ and $\sigma=1$, respectively, and the 
universal kernel $\gamma_{\,\rm u}^{}$ is reciprocity respecting (RR), i.e., 
invariant under replacement $N \to (1\!-\!N)$.
Eq.~(\ref{Stune}) implies that the non-RR parts and the spacelike$/$timelike 
difference are inherited from lower orders. Hence `only' $\gamma_{\,\rm u}^{}$,
which includes $2^{\:\!w-1\!}$ RR (combinations of) harmonic sums of weight 
$w$, needs to be determined at four loops. 

\vspace*{1mm}
Present information, given by the even-$N$ (odd-$N$) values $N\leq 16\;$(15) 
of $\gamma_{\,\rm ns}^{\,+(3)}(N)$ ($\gamma_{\,\rm ns}^{\,-(3)}(N)$) and 
endpoint constraints (see below), is insufficient to determine the $n=3$
coefficients in eq.~(\ref{gForm}).
However, $\,\gamma_{\,\rm ns}^{\,+\,} = \gamma_{\,\rm ns}^{\,-}$ in the 
large-$n_c$ limit, hence the known even-$N$ {\it and} odd-$N$ values can be 
used. 
Moreover, alternating sums do not contribute to $\,\gamma_{\,\rm ns}^{\,\pm}$ 
in this limit, leaving 1, 1, 2, 3, 5, 8, 13 $\,=\,$ Fibonacci$(w)$ RR sums at 
weight $w = 1,\ldots, 7$ and a total of 87 basis functions for $n=3$ in
eq.~(\ref{gForm}).

\vspace*{1mm}
Large$\:\!$-$N$ and small-$x$ limits provide more than 40 constraints on their
coefficients. At large-$N$, the non-singlet anomalous dimensions have the form
\cite{Korch89,ABall,DMS05}
\beq
\label{gNinf} 
  \gamma_{\,\rm ns}^{\:(n-1)}(N) \;=\;
    A_n \ln \widetilde{N} - B_n
  + N^{\:\!-1} \{ C_n \ln \widetilde{N}  
  - \widetilde{D}_n + {\textstyle\frac{1}{2}}\, A_n \}
  + O( N^{\:\!-2} )
\eeq
with $\ln \widetilde{N} \,\equiv\, \ln N + \gamma_{\,\rm e}$, where 
$\gamma_{\,\rm e}$ denotes the Euler-Mascheroni constant.
$C_n$ and $\widetilde{D}_n$ are given by
\beq
\label{CDofAB}
  C(\ars) \;=\; \left( \:\! A(\ars) \:\! \right)^{2^{}}
\;\; , \quad
  \widetilde{D}(\ars) \;=\; A(\ars) \cdot ( B(\ars)_{} - \beta(\ars)/\ars )
\:\: ,
\eeq
in terms of lower-order information on the cusp anomalous dimension
 $A(\ars) = A_1 \ars + A_2\, \ar(2) + \ldots$
and the quantity
 $B(\ars) = B_1 \ars + B_2\, \ar(2) + \ldots$
sometimes called the virtual anomalous dimension.

\vspace*{1mm}
The resummation of small-$x$ double logarithms \cite{KL82,BV95,avLL2012,DKV18}
provides the four-loop coefficients of  $\,x^{\,a} \ln^{b\!}x\,$ at 
$4 \leq b \leq 6$ and all $a$ in the large-$\nc$ limit
(in full QCD, this holds only at even $a$ for $P_{\,\rm ns}^{\,+}(x)$ and
odd $a$ for $P_{\,\rm ns}^{\,-}(x)$).
Moreover, a relation leading to a single-logarithmic resummation at $a=0\,$,
\beq
\label{SLxto0}
  \gamma_{\,\rm ns}^{\,+}(N) \!\cdot\! \left( \,\gamma_{\,\rm ns}^{\,+}(N)
  + N - \beta(\ars) / \ars \right) \,=\, O(1)
\; ,
\eeq
has been conjectured in ref.~\cite{VelizSL}. As far as it can be checked so 
far, this relation is found to be correct except for terms with $\z2 = \pi^2/6$
that vanish in the large-$n_c$ limit.

\vspace*{1mm}
Taking into account all the above information, it is possible to set up 
systems of Diophantine equations for the coefficients $c_{00\vec{w}}$,
$c_{ak\vec{w}}$ of $\gamma_{\,\rm ns}^{\,\pm(3)}(N)$ in the large-$n_c$ limit
that can be solved using the moments $ 1 \leq N \leq 18$, leaving the results
of the diagram calculation at $N \!=\! 19, 20$ as checks.
 
\vspace*{-1mm}
\section{All-$N$ anomalous dimension in the large-$n_c$ limit}
\vspace*{-1mm}

\nin
The exact expressions for the new $\nfz$ and $\nfo$ parts cannot be shown here 
due to their length, they can be found in eq.~(3.6) and (3.7) of 
ref.~\cite{MRUVV}.  For the $\nfs$ and $\nft$ terms see ref.~\cite{DRUVV}.
The resulting large-$N$ coefficients $A_{L,4}$ and $B_{L,4}$ 
-- the subscript $L$ indicates the large-$n_c$ limit -- are found to be
\bea
\label{A4Lnc}
  A_{L,4} &\!=\!& 
         \cf \* \nct \* \!\left( \,
            \frct{84278}{81}\,
          - \frct{88832}{81}\,\*\z2
          + \frct{20992}{27}\,\*\z3
          + 1804\,\*\z4
          - \frct{352}{3}\,\*\z2\*\z3
          - 352\,\*\z5
          - 32\,\*\zts
          - 876\,\*\z6 
\!
          \right)
\nonumber \\[1mm] && \mbox{\hspn}
       - \,\cf\*\ncs\*\nf \* \!\left( \,
            \frct{39883}{81}
          - \frct{26692}{81}\,\*\z2
          + \frct{16252}{27}\,\*\z3
          + \frct{440}{3}\,\*\z4
          - \frct{256}{3}\,\*\z2\*\z3
          - 224\,\*\z5
\!
          \right)
\nonumber \\[1mm] && \mbox{\hspn}
        + \,\cf\*\nc\*\nfs \* \!\left( \,
            \frct{2119}{81}
          - \frct{608}{81}\,\*\z2
          + \frct{1280}{27}\,\*\z3
          - \frct{64}{3}\,\*\z4
\!
          \right)
        \,-\, \cf\*\nft \* \!\left( \,
            \frct{32}{81}
          - \frct{64}{27}\,\*\z3
\!
          \right)
\eea
and
\bea
\label{B4Lnc}
 B_{L,4} &\!=\!& 
         \cf \* \nct \* \left(
          - \frct{1379569}{5184}
          + \frct{24211}{27} \,\* \z2
          - \frct{9803}{162} \,\* \z3
          - \frct{9382}{9} \,\* \z4
          + \frct{838}{9} \,\* \z2 \* \z3
          + 1002 \* \z5
          + \frct{16}{3} \,\* \zts
 \right. \nonumber \\ && \left. \mbox{\hspp\hspp} \phantom{\frct{1}{2}}
          + 135 \* \z6
          - 80 \* \z2 \* \z5
          + 32 \* \z3 \* \z4
          - 560 \* \z7
          \right)
\nonumber \\[0.5mm] && \mbox{\hspn}
       +\, \cf\*\ncs\*\,\nf \* \left(
            \frct{353}{3}
          - \frct{85175}{162} \,\* \z2
          - \frct{137}{9} \,\* \z3
          + \frct{16186}{27} \,\* \z4
          - \frct{584}{9} \,\* \z2 \* \z3
          - \frct{248}{3} \,\* \z5
          - \frct{16}{3} \,\* \zts
          - 144 \* \z6
\!
          \right)
\nonumber \\[1mm] && \mbox{\hspn}
       -\, \cf\*\nc\*\,\nfs \* \left(
            \frct{127}{18}
          - \frct{5036}{81} \,\* \z2
          + \frct{932}{27} \,\* \z3
          + \frct{1292}{27} \,\* \z4
          - \frct{160}{9} \,\* \z2 \* \z3
          - \frct{32}{3} \,\* \z5
          \right)
\nonumber \\[1mm] && \mbox{\hspn}
       -\, \cf\*\nft \* \left(
            \frct{131}{81}
          - \frct{32}{81} \,\* \z2
          - \frct{304}{81} \,\* \z3
          + \frct{32}{27} \,\* \z4
          \right)
\; . 
\eea
The agreement of the four-loop cusp anomalous dimension (\ref{A4Lnc}) with 
the result obtained from the large-$n_c$ photon-quark form factor 
\cite{cusp4Lnc1,cusp4Lnc2} provides a further non-trivial check of the 
determination of the all-$N$ expressions from the moments at $N \leq 18$, 
and hence also of the relations (\ref{gForm}) -- (\ref{SLxto0}). 
 
\vspace*{1mm}
The maximum-weight $\zts$ and $\z6$ parts of eq.~(\ref{A4Lnc}) also agree 
with the result obtained in planar ${\cal N} \!= 4$ maximally supersymmetric 
Yang-Mills theory (MSYM) obtained before in ref.~\cite{cusp4N=4}. 
There is no such direct connection between the four-loop virtual anomalous
dimension (\ref{B4Lnc}) and its counterparts in planar ${\cal N} \!= 4$ MSYM; 
see ref.~\cite{Dixon17} where the maximum-weight part of eq.~(\ref{B4Lnc}) has 
been employed to derive the four-loop collinear anomalous dimension in planar 
${\cal N} \!= 4$ MSYM.

\vspace*{1mm}
The all-$N$ large-$n_c$ limit of $\gamma_{\,\rm ns}^{\,\pm(3)}(N)$ is compared 
in fig.~1 with the integer-$N$ QCD results at $N \leq 16$. As illustrated in 
the left panel, the former are a decent approximation to the latter for the 
individual $n_{\! f}^{\,k}$ contributions. However, as shown in the right
panel, there are considerable cancellations between the these contributions. 
These cancellations are most pronounced for the physically relevant number of 
$\nf = 5$ light quark flavours outside the large-$N\,/\,$large-$x$ region. 
Hence the \mbox{large-$n_c$} suppressed contributions -- indicated by the 
subscript $N$ below -- need to be taken into account in phenomenological 
N$^3$LO analyses.

\begin{figure}[t]
\vspace{-1mm}
\centerline{\epsfig{file=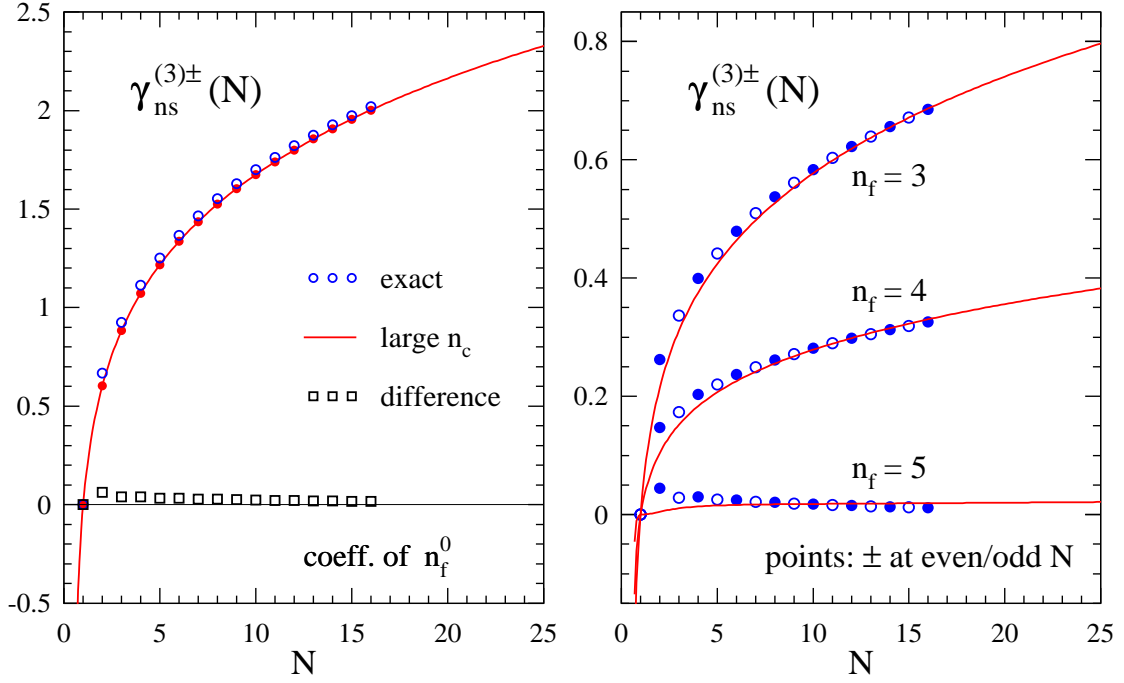,width=15.0cm,angle=0}}
 \caption{ 
 The large-$n_c$ limit of the four-loop anomalous dimensions
 $\,\gamma_{\,\rm ns}^{\,\pm(3)}(N)$ (lines) compared to the QCD 
 results for $\gamma_{\,\rm ns}^{\,+(3)}(N)$ at even $N$ and 
 $\gamma_{\,\rm ns}^{\,-(3)}(N)$ at odd $N$ (points). 
 Left: the $\nf$-independent
 contributions. Right: the results for physically relevant values of $\nf$.
 The values have been converted to an expansion in $\als$.
 }
\vspace{-1mm}
\end{figure}

\section{$x\:\!$-$\:\!$space approximations of the large-$n_c$ suppressed parts}
\vspace*{-1mm}

\nin
With eight integer-$N$ moments known for both $P_{\,\rm ns}^{\,+(3)}(x)$ and 
$P_{\,\rm ns}^{\,-(3)}(x) $ and the large-$x$ and 
\mbox{small-$x$} knowledge discussed
in section 2, it is possible to construct approximate $x$-space expressions 
which are analogous to (but more accurate than) those used before 2004 at 
N$^2$LO, see refs.~\cite{NV1,NV2,NV4,MVV1}. 
For this purpose an ansatz consisting of
\begin{itemize}
\item the two large-$x$ parameters $A_4$ and $B_4$ in eq.~(\ref{gNinf}),
      \\[-8mm]
\item two of three suppressed large-$x$ logs
      $\:\!\xm1 \ln^{\:\!k\!}\xm1$, $k = 1,2,3$,
      \\[-8mm]
\item one of ten two-parameter polynomials in $x$ that vanish for
      $x \!\to\! 1$,
      \\[-8mm]
\item two of the three unknown small-$x$ logarithms
      $\:\!\ln^{\:\!k\!}x$, $k = 1,2,3$
\end{itemize}
is built for the large-$n_c$ suppressed $\nfz\:\!$ and $\nfo$ parts 
$P_{{\rm N},0/1}^{\,+(3)}$ of $P_{\,\rm ns}^{\,+(3)}(x)\:\!$. 
This results in 90 trial functions, the parameters of which can be fixed from 
the eight available moments. Of these functions, two representatives $A$ and 
$B$ are then chosen that indicate the remaining uncertainty, see fig.~2.

\vspace*{1mm}
This                non-rigorous procedure can be checked by comparing the
same treatment for the large-$n_c$ parts to our exact results. 
Moreover, the trial functions lead to very similar values for the next moment, 
e.g., $N=18$ for $P_{\,\rm ns}^{\,+(3)}$.
The residual uncertainty at this $N$-value is a consequence of the width of
the band at large $x$, which in turn is correlated with the uncertainties at 
smaller $x$.
If the spread of the result $A$ and $B$ would underestimate the true remaining
uncertainties, then a comparison with an additional analytic result at this
next value of $N$ should reveal a discrepancy.
We were able to extend the diagram computations of the $\nfo$ parts of 
$P_{\,\rm ns}^{\,(3)+}(x)$ to $N=18$ and find
\beq
  P_{\:\!{\rm N},1}^{\,+(3)}(N\!=\!18) \:=\:
            195.8888792_{\:\!B} 
    \; < \; 195.8888857..._{\,\rm exact} 
    \; < \; 195.8888968_A
\:\: .
\eeq
A similar check for $P_{\:\!{\rm N},0}^{\,+(3)}$ has been carried out by 
deriving a less accurate approximation using only seven moments and comparing 
the results to the now unused value at $N=16$.

\vspace*{1mm}
The case of $P_{\,\rm ns}^{\,-(3)}(x)$ has been treated in the same manner,
but taking into account that only its leading small-$x$ logarithm is
known up to now \cite{BV95}.
See ref.~\cite{MRUVV} for the (large-$N$ suppressed) additional 
$d^{\,abc}d_{abc}$ contribution $P_{\,\rm ns}^{\:{\rm s}(3)}(x)$ to the
splitting function for the total valence quark PDF.
  
\begin{figure}[t]
\vspace{-1mm}
\centerline{\hspace*{1mm}\epsfig{file=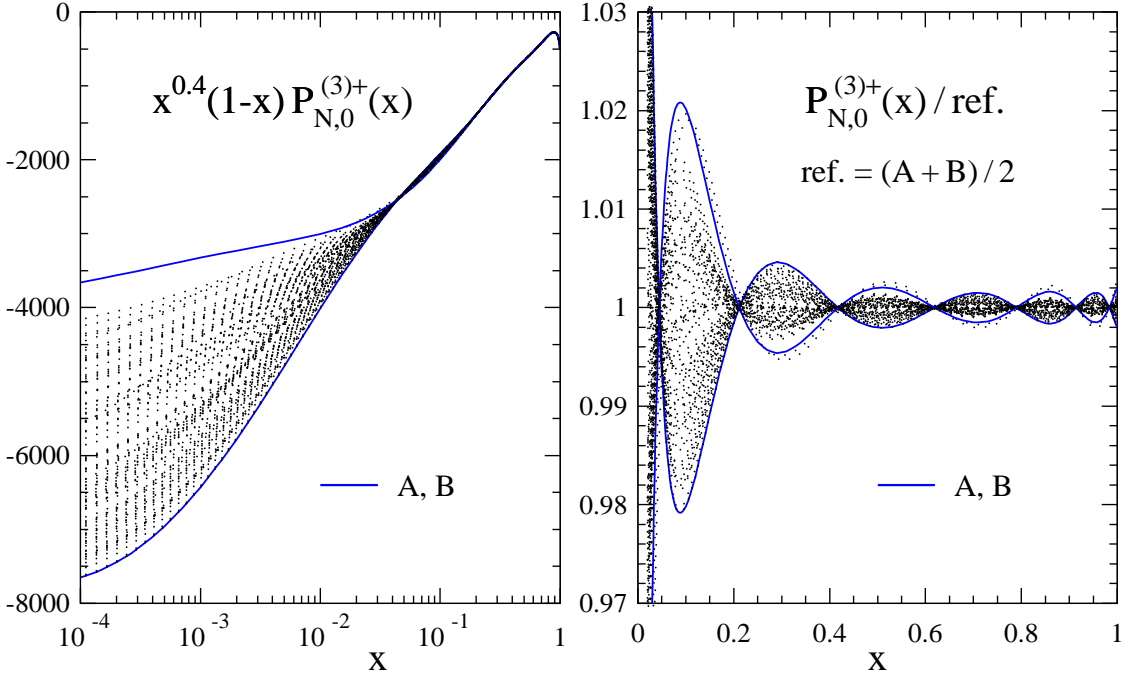,width=15.0cm,angle=0}}
\vspace{-3mm}
\caption{ 
 About 90 trial functions for the $\nf$-independent contribution to the
 large-$n_c$ suppressed part of splitting function $P_{\,\rm ns}^{\,+(3)}(x)$,
 multiplied by $x^{\,0.4}\xm1_{}$.
 The two functions chosen to represent the remaining uncertainty are 
 denoted by $A$ and $B$ and shown by solid (blue) lines.
 Due to the factor $\xm1_{}$ the contribution $A_{N,4}$ to the four-loop 
 cusp anomalous dimension can be read off at $x\!=\!1$.
 }
\vspace{-3mm}
\end{figure}

\vspace*{-1mm}
\section{Numerical results for the cusp and virtual anomalous dimensions}
\vspace*{-1mm}

\nin
Combining the exact large-$n_c$ results, the approximations for the remaining
$\nfz$ and $\nfo$ contributions and the complete high-$\nf$ contributions of 
ref.~\cite{DRUVV}, the four-loop cusp anomalous dimension for QCD with $\nf$ 
quark flavours are given by
\beq
   A_4 \;=\;   20702(2)       - 5171.9(2) \,\nf
             + 195.5772\,\nfs + 3.272344 \,\nft
\; ,
\eeq
where the numbers in brackets represent a conservative estimate of the 
remaining uncertainty. The conversion of this result to an expansion in 
powers of $\als$ leads to
\bea
\label{Aqnum}
  A_q(\als,\nf\!=\!3) &=& 0.42441\,\als \,
  ( 1 \,+\, 0.72657\,\als \,+\, 0.73405\,\as(2) \,+\, 0.6647(2)\,\as(3) 
    + \ldots\, )
\; , \nn \\
  A_q(\als,\nf\!=\!4) &=& 0.42441\,\als \,
  ( 1 \,+\, 0.63815\,\als \,+\, 0.50998\,\as(2) \,+\, 0.3168(2)\,\as(3) 
    + \ldots\, )
\; , \nn \\
  A_q(\als,\nf\!=\!5) &=& 0.42441\,\als \,
  ( 1 \,+\, 0.54973\,\als \,+\, 0.28403\,\as(2) \,+\, 0.0133(2)\,\as(3) 
    + \ldots\, )
\; .
\eea
The corresponding results for the virtual anomalous dimension, i.e., the
coefficient of $\delta (1\!-\!x)$ show a similarly benign expansion with 
\beq
   B_4 \;=\;   23393(10)      - 5551(1) \,\nf
             + 193.8554\,\nfs + 3.014982\,\nft
\eeq
 
\vspace*{-3mm}
\nin
and
\bea
\label{Bqnum}
  B_q(\als,\nf\!=\!3) &=& 0.31831\,\als \,
  ( 1 \,+\, 0.99712\,\als \,+\, 1.24116\,\as(2) \,+\, 1.0791(13)\,\as(3) 
    + \ldots\, )
\; , \nn \\
  B_q(\als,\nf\!=\!4) &=& 0.31831\,\als \:
  ( 1 \,+\, 0.87192\,\als \,+\, 0.97833\,\as(2) \,+\, 0.5649(13)\,\as(3) 
    + \ldots\, )
\; , \nn \\
  B_q(\als,\nf\!=\!5) &=& 0.31831\,\als\:
  ( 1 \,+\, 1.74672\,\als \,+\, 0.71907\,\as(2) \,+\, 0.1085(13)\,\as(3) 
    + \ldots\, )
\; .
\eea
Due to constraints by large-$N$ moments, the errors of $A_4$ and $B_4$ are 
fully correlated. The accuracy in eqs.~(\ref{Aqnum}) and (\ref{Bqnum}) should 
be amply sufficient for phenomenological applications.

\vspace*{1mm}
By repeating the approximation procedure in section 5 for individual colour 
factors, it is possible to obtain corresponding approximate coefficients for 
$A_4$ and $B_4$ which can be summarized~as 
(for a table of the relevant group invariants see, e.g., appendix C of 
ref.~\cite{RUVV17})
\begin{center}
  \renewcommand{\arraystretch}{1.2}
  \begin{tabular}{MMM}
   \mbox{ }      &    A_4                      &  B_4     \\[1pt]
   \hline\\[-5mm]
    \cff         &     0                       &  ~~~~197. \,\pm\, ~3. \\
    \cft\, \ca   &     0                       &   \;-687. \,\pm\, 10. \\
    \cfs \cas    &     0                       &  ~~~1219. \,\pm\, 12. \\
    \cf \cat     & \phantom{-}  610.3 \,\pm\, 0.3 & ~~~295.6 \,\pm\,2.4\\
    d_R^{\,abcd}d_A^{\,abcd}/N_{\!R}^{}
                 &    -507.5 \,\pm\, 6.0       &   \;-996. \,\pm\, 45. \\[0.2mm]
\hline\\[-5mm]
    \nf\, \cft   & -31.00 \pm 0.4     & \phantom{-0} 81.4 \pm 2.2      \\
  \nf\,\cfs\ca   & \phantom{-} 38.75 \,\pm\, 0.2  & -455.7 \,\pm\, 1.1 \\
  \nf\,\cf\cas   &  -440.65 \,\pm\, 0.2~~      &    -274.4 \,\pm\, 1.1 \\
    \nf\,d_R^{\,abcd}d_R^{\,abcd}/N_{\!R}^{}
                 &  -123.90  \,\pm\, 0.2~~     &    -143.5 \,\pm\, 1.2 \\[0.2mm]
\hline\\[-5mm]
    \nfs\,\cfs   &   -21.31439         & -5.775288 \\[-0.4mm]
  \nfs\,\cf\ca   & \phantom{-}58.36737 & \phantom{-}51.03056 \\[-0.4mm]
    \nft\,\cf    & \phantom{-}2.454258 & \phantom{-}2.261237 \\[0.2mm]
  \hline
  \end{tabular}
\end{center}

\vspace*{1mm}
\nin
where the exactly known $\nfs$ and $\nft$ coefficients have been included for
completeness. 
Due to the constraint provided by the exact large-$n_c$ limit, the errors 
in this table are highly correlated; for numerical applications in QCD 
eqs.~(\ref{Aqnum}) and (\ref{Bqnum}) should be used instead. 
The above results show that both quartic group invariants definitely contribute
to the four-loop cusp anomalous dimension, for this issue see also refs.~%
\cite{GrozinHKM15,BoelsHY17a,GrozinHS17,BoelsHY17b} and references therein.
This implies that the so-called Casimir scaling between the quark and gluon
cases, $\,A_q \,=\, \cf/\ca\, A_g\,$, does not hold beyond three loops.

\section{N$^3$LO corrections to the evolution of non-singlet PDFs}
\vspace*{-1mm}

\nin
The effect of the fourth-order contributions on the  evolution of the 
non-singlet PDFs can be illustrated by considering the logarithmic derivatives 
of the respective combinations of quark PDFs with respect to the factorization 
scale,
$\,\dot{q}_{\rm ns}^{\:i} \equiv d\ln q_{\rm ns}^{\:i}/ d\ln \mu_{\!f}^{2}$,
at a suitably chosen reference point. \linebreak

\pagebreak
\nin
As in ref.~\cite{P2ns}, we choose the schematic, order-independent initial
conditions
\beq
\label{Input}
  xq_{\rm ns}^{\,\pm,\rm v}(x,\mu_{0}^{\,2}) \; = \; x^{\, 0.5} (1-x)^3
\quad \mbox{ and } \quad
  \als (\mu_{0}^{\,2}) \:=\: 0.2
\; .
\eeq
For $\als (M_Z^{\, 2}) = 0.114 \ldots 0.120$ this value for $\als$ corresponds 
to $\mu_{0}^{\,2} \,\simeq\, 25\ldots 50$ GeV$^2$ beyond the leading order, 
a scale range typical for DIS at fixed-target experiments and at the {\it ep} 
collider HERA.

\vspace*{1mm}
The new N$^3$LO corrections to $\,\dot{q}_{\rm ns}^{\, i}$ are generally small,
hence they are illustrated in fig.~3 by comparing their relative effect to that
of the N$^2$LO contributions for the standard identification $\mu_r = \mu_{\!f}
\equiv \mu$ of the renormalization scale with the factorization scale. 
Except close to the sign change of the scaling violations at $x \simeq 0.07$,
the relative N$^3$LO effects are (well) below 1\% for the flavour-differences 
$q_{\rm ns}^{\, +}$ and $q_{\rm ns}^{\, -}$ (left and middle panel).
The N$^2$LO and N$^3$LO corrections are larger for the valence distribution
$q_{\rm ns}^{\,\rm v}$ at $x < 0.07$ due to the effect of the 
$d^{\:\!abc}d_{abc}$ `sea' contribution $P_{\,\rm ns}^{\,\rm s}(x)$, note the 
different scale of the right panel in fig.~3.
Also in this case the N$^3$LO evolution represents a clear improvement, and
the relative four-loop corrections are below 2\%.

\vspace*{1mm}
The remaining uncertainty due to the approximate character of the four-loop
splitting functions beyond the large-$n_c$ limit is indicated by the difference
between the solid and dotted (red) curves in fig.~3 and fig.~4 below.
Due to the small size of the four-loop contributions and the `$x$-averaging' 
effect of the Mellin convolution, 
\beq
\label{Mconv}
  [ \,P_{\,\rm ns} \otimes q_{\rm ns}^{} ](x) \;=\;
  \int_x^1  \frct{dy}{y} \: P_{\,\rm ns} (y)\:
  q_{\rm ns}^{}\Big(\,\frct{x}{y}\,\Big)
\; ,
\eeq
the results of section 4 are safely applicable to lower values of $x$ than 
one might expect from fig.~2.

\vspace*{1mm}
The stability of the NLO, N$^2$LO and N$^3$LO results under variation of the 
renormalization scale over the range $\frac{1}{8}\,\mu_{\!f}^{\,2} \,\leq\, 
\mu_r^{\,2} \,\leq\, 8\:\!\mu_{\!f}^{\,2}$ is illustrated in fig.~4 at typical
values of $x$. 
Except close to the sign change of $\,\dot{q}_{\rm ns}^{\, +}$,
the variation is well below 1\% for the conventional interval
$\frac{1}{2}\:\mu_{\!f}^{} $ $\leq$ $\mu_r^{}$ $\leq$ $2\:\! \mu_{\!f\,}^{}$.

\begin{figure}[b]
\vspace{-5mm}
\centerline{\hspace*{0mm}\epsfig{file=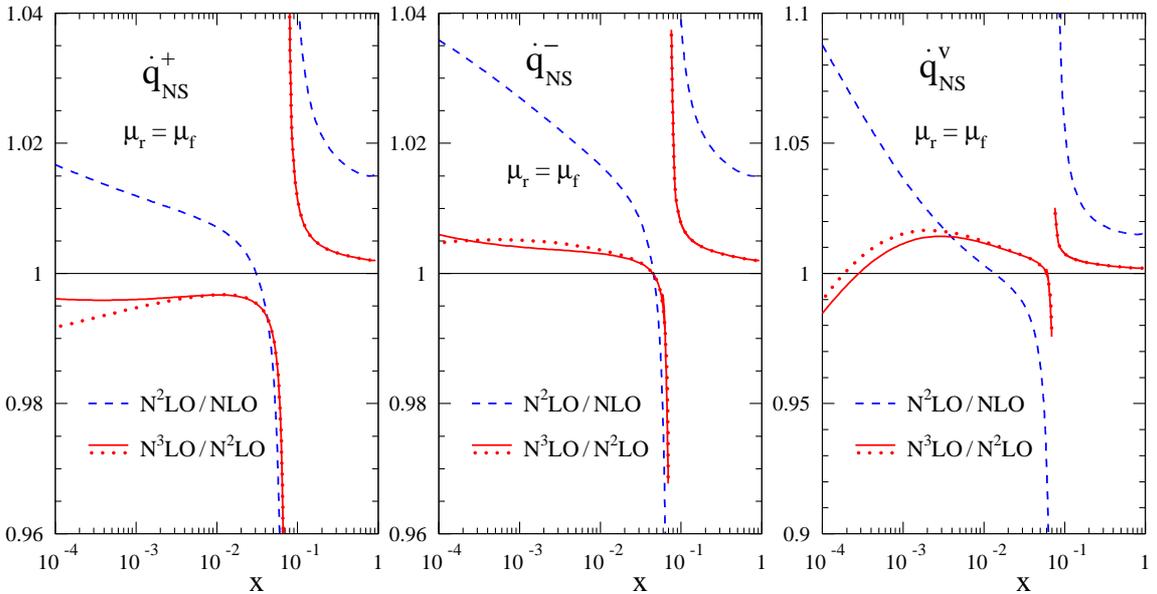,width=15.4cm,angle=0}}
\vspace{-2mm}
\caption{
 The relative N$^2$LO and N$^3$LO corrections to the logarithmic scale 
 derivative of the non-singlet combinations $q_{\,\rm ns}^{\,a}$ of quark
 PDFs for the schematic order-independent input (\protect\ref{Input}) for 
 $\nf = 4$ at $\mu_r = \mu_{\!f\,}$.
 }
\vspace*{-1mm}
\end{figure}
 
\begin{figure}[t]
\vspace{-1mm}
\centerline{\hspace*{0mm}\epsfig{file=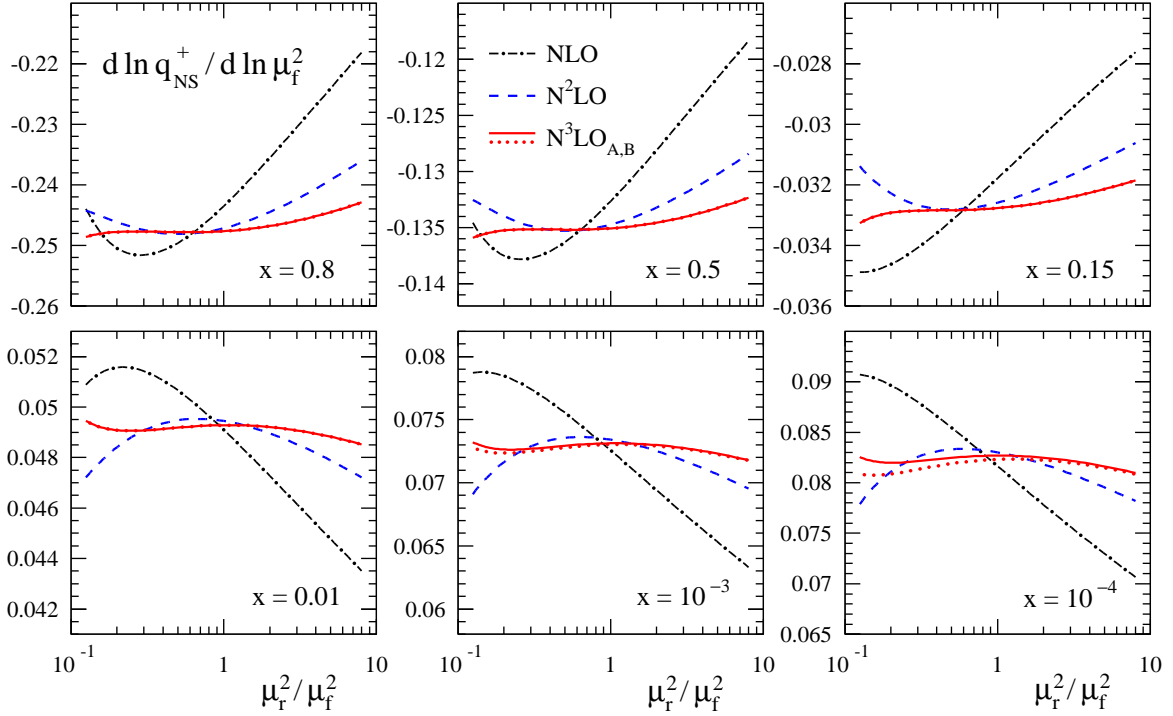,width=15.4cm,angle=0}}
\vspace{-3mm}
\caption{
 The dependence of the NLO, N$^2$LO and N$^3$LO results for
 $\dot{q}_ {\rm ns}^{\, +} \:\equiv\: d\ln q_{\rm ns}^{\, +}/ d\ln\mu_{\!f}^2$
 on~the renormalization scale $\mu_r^{}$ at six typical values of $x$ for the 
 initial conditions (\protect\ref{Input}) and $\nf = 4$ flavours.
 The remaining uncertainty of the four-loop splitting function 
 $P_{\,\rm ns}^{\,+(3)}(x)$ leads to the difference of the solid and 
 dotted~curves.
 }
\vspace*{-1mm}
\end{figure}

\vspace*{-1mm}
\section{Summary and Outlook}
\vspace*{-1mm}
 
\nin
The splitting functions for the non-singlet combinations of quark PDFs
have been addressed at the fourth-order (N$^3$LO) of perturbative QCD. 
The quantities $P_{\,\rm ns}^{\,\pm (3)}$ are now known exactly in the limit 
of a large number of colours $n_c$. Present results for the large-$n_c$ 
suppressed contributions with $\nfz$ and $\nfo$ are still approximate, but 
sufficiently accurate for phenomenological applications in deep-inelastic 
scattering and collider physics. {\sc Form} and {\sc Fortran} files of 
these results can be obtained by downloading the source of ref.~\cite{MRUVV} 
from {\tt arXiv.org}.

\vspace*{1mm}
It would be desirable, mostly for theoretical purposes, to obtain also the
analytic forms $\nfz$ and $\nfo$ parts of $P_{\,\rm ns}^{\,\pm (3)}$. So far,
only their contributions proportional to the values $\z4$ and $\z5$ of the
Riemann $\zeta$-function have been completely determined, together with the
(unpublished) $\z3$ part of the $\nfo$ contributions.
The $\z4$ parts are particularly simple; in fact, it turns out that they 
(and other $\pi^2$ terms) can be predicted via physical evolution kernels 
from lower-order quantities,~see~refs.~\cite{JM-no-pi2,DV-no-pi2}.

\vspace*{1mm}
The $\z5$ part of $P_{\,\rm ns}^{\,\pm (3)}$, presented in appendix D of
ref.~\cite{MRUVV}, includes a (non large-$n_c$) contribution 
\beq
\label{z5S1sq}
  -\,\frct{128}{3}\,\left\{ 3\,\cfs\,\cas - 2\,\cf\cat + 12\:\dfFAnc \right\}
   \:5\:\!\z5\, [S_1(N)]^2
\; .
\eeq
The resulting $\ln^2 N$ large-$N$ behaviour needs to be compensated by 
non-$\zeta_5$ terms. Eq.~(\ref{z5S1sq}) looks exactly like the $\z5$-`tail' 
of the so-called wrapping correction in the anomalous dimensions in 
\mbox{$\,{\cal N}\!=4$} maximally supersymmetric Yang-Mills theory, 
see refs.~\cite{Kotikov:2007cy,Bajnok:2008qj}.

\pagebreak

Phenomenologically, of course, one rather needs corresponding results for the 
flavour-singlet splitting functions $P_{ij}^{\,(3)}(x)$, $\,i,j = q,g$.  
At present, it appears computationally too hard to obtain moments of all four functions 
beyond $N = 6$ using the method of refs.~\cite{LV1991,3loopN1,3loopN2,3loopN3}.
Therefore one will need to resort to the OPE, which offers additional
theoretical challenges in the massless flavour-singlet case, see 
refs.~\cite{Hamberg:1991qt,Collins:1994ee,Matiounine:1998ky}.
We hope to address this issue in a future publication.

\subsection*{Acknowledgements}
\vspace*{-1mm}

\nin
The research reported here has been supported by 
the {\it European Research Council}$\,$ (ERC) \mbox{Advanced} Grant 320651,
{\it HEPGAME},
the  grant ST/L000431/1 of the UK {\it Science \& Technology Facilities 
Council}$\,$ (STFC), 
and the {\it Deutsche Forschungsgemeinschaft} (DFG) grant MO~1801/1-2 
and SFB 676 project A3.
Part of our computations have been performed on a computer cluster 
in Liverpool funded by the STFC grant ST/H008837/1.

\end{document}